\pdfoutput=1

\documentclass{iopart}
\usepackage{graphicx} 

\begin{document}

\article[]{VIEWPOINT}{The net charge at interfaces between insulators}

\author{N C Bristowe$^{1,2}$, P B Littlewood$^1$ and Emilio Artacho$^2$}

\address{$^1$Theory of Condensed Matter Group,
             Cavendish Laboratory, University of Cambridge, 
             JJ Thomson Ave, Cambridge CB3 0HE, UK}

\address{$^2$Department of Earth Sciences, University of Cambridge, 
             Downing Street, Cambridge CB2 3EQ, UK}

\ead{ncb30@cam.ac.uk}
\begin{abstract}
  The issue of the net charge at insulating oxide interfaces is
shortly reviewed with the ambition of dispelling myths of 
such charges being affected by covalency and related 
charge density effects.
  For electrostatic analysis purposes, the net charge at 
such interfaces is defined by the counting of discrete electrons 
and core ion charges, and by the definition of the reference 
polarisation of the separate, unperturbed bulk materials.
  The arguments are illustrated for the case of a thin film of
LaAlO$_3$ over SrTiO$_3$ in the absence of free carriers,
for which the net charge is exactly 0.5$e$ per interface formula 
unit, if the polarisation response in both materials is referred to 
zero bulk values.
  Further consequences of the argument are extracted
for structural and chemical alterations 
of such interfaces, in which internal rearrangements are 
distinguished from extrinsic alterations (changes of 
stoichiometry, redox processes), only the latter affecting 
the interfacial net charge. 
  The arguments are reviewed alongside the proposal of
Stengel and Vanderbilt [Phys. Rev. B {\bf 80}, 241103 (2009)]
of using formal polarisation values instead of net interfacial 
charges, based on the interface theorem of Vanderbilt and 
King-Smith [Phys. Rev. B {\bf 48}, 4442 (1993)].
  Implications for non-centrosymmetric materials are 
discussed, as well as for interfaces for which the charge 
mismatch is an integer number of polarisation quanta.
\end{abstract}


\section{Introduction}

  When Ohtomo and Hwang \cite{Ohtomo2004} showed how to produce two-dimensional
electron gases in pristine insulator-insulator interfaces, they rationalised 
their discovery in terms of the electrostatics of formally charged ions.
  Namely, $-2 e$ for O, $+4 e$ for Ti, $+2 e$ for Sr, and $+3 e$ 
for Al and La, respectively, for their original work with a 
LaAlO$_3$/SrTiO$_3$ (001) interface (LAO/STO).
  With such charges the alternating (001) SrO and TiO$_2$ atomic layers 
of the perovskite structure of STO are neutral, while the LAO corresponding
ones, LaO and AlO$_2$, are charged as  $+1e$ and $-1e$
per formula unit, respectively.
  These net layer charges give rise to an imbalance at the interface giving
a net charge of $\pm 0.5 e$ per interface formula unit (the sign depending 
on whether the interface is SrO/AlO$_2$ or LaO/TiO$_2$).
  This basic argument is at the heart of all the subsequent work in what
has become a very hot topic \cite{Mannhart2010}.

  It is well known, however, that formal charges represent quite an
idealised picture of the electronic distribution in these far from
ideally ionic insulators, in which covalency effects (notably in 
Al-O and Ti-O bonds) are not at all negligible.
  Indeed, many papers with quantitative modelling of the electrostatics 
allow for covalency effects, and thus a reduction of the interface 
effective charge. 

\section{Basic argument}

  The intention of this note is to dispel covalency myths when dealing
with the electrostatics of these systems (as reviewed in e.g. 
Ref.~\cite{Goniakowski2008}): if the bulk polarisation of 
both materials is taken as zero, the effective charge of these pristine 
band-insulating interfaces is exactly what obtained from counting formal 
charges as done in the original paper~\cite{Ohtomo2004}, absolutely 
irrespective of ionicity or covalency considerations.
  It is exactly $0.5 e$ per formula unit for LAO/STO (in the absence of 
free carriers trapped at the interface, which will be assumed here unless 
explicitly stated).  
  
  The key to the argument is that what matters is charge counting, 
not the shape of the charge density.
  This is not new, it has already been said for these systems (see 
Ref.~\cite{Bristowe2009}, also implicit in Ref.~\cite{Stengel2009}), 
but was already understood for polar/non-polar semiconductor 
interfaces \cite{Baraff1977} and formalised in Ref.~\cite{Vanderbilt1993}.
  It is the old argument for understanding doping in semiconductors.
  Take phosphorous as a donor in bulk silicon.
  Its $+5$ core is surrounded by the same set of bonds that surrounded the
Si $+4$ core it substitutes.
  These bonds can be described by electron pairs in localised 
Wannier functions (e.g. maximally localised Wannier functions 
\cite{Marzari1997}).
  These functions may be deformed as compared to what they would 
be in the pristine material, but the number of such functions remains, since 
they relate to the number of bands in the valence band.
  Take a large enough region around the dopant.
  The number of electrons in that region described by the full valence band 
is identical to the number in the same region without the dopant substitution.
  Since the core is $+5$ instead of $+4$, the net charge is precisely $+1e$.
  The fifth electron of P is then shallowly bound to that charge,
in a hydrogenic Rydberg state below the conduction band, defined 
by the effective mass of the relevant conduction-band minimum, 
the dielectric constant of the material (which describes the deformation 
of the bonds around the dopant), and $Z=+1$. 

  In the LAO/STO system the interface is no longer a point defect
and has a different material at each side. 
  However, the valence bands of both LAO and STO give rise to
four Wannier functions per O atom.
  This does not mean that they are on O in an ideal ionic form, in 
fact they are very much deformed towards the Ti or Al nuclei, 
as polar covalent bonds. 
  It is just a counting consideration, very much as that of 
there being two bonds per atom in a Si crystal.
  Counting core ionic charges and the valence electrons that way gives
exactly the same charges as the formal charges used by Ohtomo and 
Hwang~\cite{Ohtomo2004}.
  As long as these Wannier functions localize over length scales smaller
than the relevant system size (the LAO film thickness in the case at hand),
using ``polarisation-free" unit cells~\cite{Goniakowski2008} allows 
to obtain the net interface charging of 0.5$e$ per formula unit.

\section{Formal discussion}

  The need to resorting to such ``polarisation-free cells", however, reflects
the fact that the definition of the net charge, albeit basic, is subtle.
  Let us review the argument a bit more formally.
  In the ``macroscopic" electrostatics we are interested in (length scales 
larger than atomic; extended to microscopics in Refs.
\cite{Baldereschi1988,Wu2008}) electric fields and charges relate 
through Gauss's law, $\vec \nabla \cdot \vec D = \rho$ (MKSA units), i.e., the 
divergence of the electric displacement field equals the volume density 
of free charges (free meaning beyond the bound charges of the dielectric). 
  The displacement combines the electric field $\vec {\cal E}$ and the 
material's polarisation $\vec P$ through $\vec D = \epsilon_0 \vec {\cal E} 
+ \vec P$, being $\epsilon_0$ the dielectric permittivity of vacuum.
  For an interface, Gauss's law becomes $D_{z}^{L} - D_{z}^{R} = \sigma$, where
$z$ is taken as the direction normal to the interface, $L$ and $R$ indicate the 
materials at either side of the interface (left and right), and $\sigma$ stands
for the area density of free charges associated to the interface.
  Thus, $\epsilon_0 ( {\cal E}_{z}^{L}- {\cal E}_{z}^{R}) = \sigma - 
(P_{z}^{L} - P_{z}^{R})$, or $\epsilon_0  \Delta {\cal E}_z = \sigma - \Delta P_z$.
  In this work the charge per unit area of $P$ and $\sigma$ 
are expressed as electron charges per interface formula unit, $e/$f.u.
  The interface area associated to a STO perovskite f.u. is $\sim 16$ \AA$^2$.

  The subtlety in the problem arises from the polarisation definition, 
and, related, what is meant by ``free charges".
  Remember that we consider the interface with no free carriers,
and thus no free charges in a strict sense. 
  The net charge we are discussing plays the same role,
however, given that it is not described by polarisation.
  It has been called compositional charge \cite{Murray2009}.
  The value of the polarisation in each material depends 
on the electric field acting in it.
  Such polarisation response to the electric field in
the structure is not discussed here, it is explicitly considered 
in the electrostatic modelling of the interface system, either with an explicit 
$P$ or by a dielectric constant (see Ref.~\cite{Bristowe2009} for an 
example in this context).
  The key is the reference polarisation of the separated, 
unperturbed bulk materials (zero of polarisation), since a change 
in that reference changes $\sigma$ by the same amount.
  There are two aspects to this.
  
\begin{figure}
\begin{center}
\includegraphics[width=4.5in]{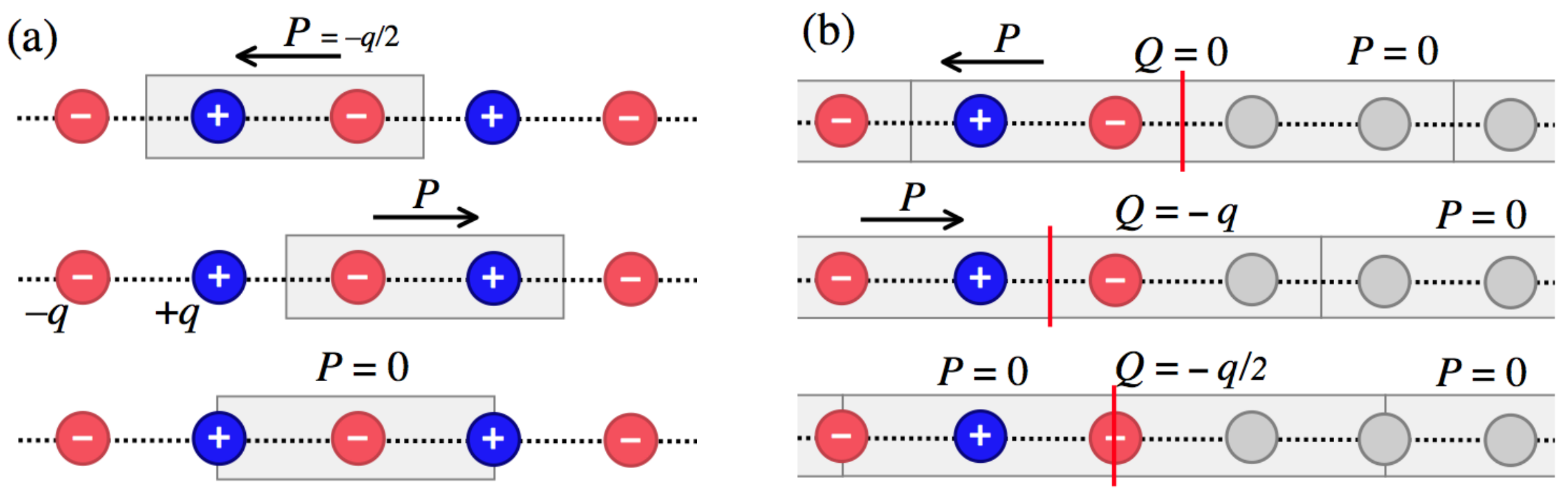}
\caption{(a) Polarisation $P$ values in a one-dimensional 
system of alternating $\pm q$ charges. As dipole moment per unit length,
$P=q (a/2)/a = q/2= P_0/2$, where $P_0$ is a quantum of polarisation
(in the real system $q=e$).
(b) Interface charge $Q$ for an interface of the previous chain with one
of neutral charges. The different values of $Q$ are associated to the 
definition of $P$.}
\end{center}
\end{figure}
  
  ($i$) The usual experimental determinations of polarisation in a material
give the change in polarisation between two states, but has no
access to an absolute value for each state separately.
  Centrosymmetric materials (as LAO and STO) offer a natural definition 
of zero polarisation, thus allowing the definition of an {\it effective} 
polarisation, as the change in polarisation when the structure deviates 
from the centrosymmetric one (we follow the name convention of 
Ref.~\cite{Stengel2009}).
  Switchable non-centrosymmetric solids (ferroelectrics) also have a 
natural zero of polarisation.
  Other non-centrosymmetric cases do not, and are discussed below.
 
  ($ii$) The determination of polarisation from theory is not straightforward
either.
  In a system of classical point charges, the polarisation of the bulk material
can be obtained as the dipole moment of the unit cell divided by the
unit-cell volume.
  This is ambiguous, however, since it depends on how the unit cell is defined.
  Take for instance a linear chain of equidistant, alternating $+q$ and $-q$ 
point charges (see Figure 1).  
  If you take the unit cell as having $+q$ on the left and $-q$ on the right,
you get a different polarisation than if you choose it the other way around 
(actually, you get the same value with opposite sign).
  You can also choose your cell by taking $+q$ in the centre, and one half
of the $-q$ charge at either end of the unit cell.
  This one achieves zero polarisation by splitting the charge.  
  If an interface is made to a similar chain of neutral particles (Figure 1), 
the following descriptions are equivalent: 
  ($i$) $\Delta P=-q/2$ 
and net interface charge $Q=0$; 
  ($ii$) $\Delta P=+q/2$ and $Q=-q$; or
  ($iii$) $\Delta P=0$ and $Q=-q/2$.
  In all cases, however, the change in electric field remains the same, 
since $\epsilon_0 \Delta {\cal E} = Q - \Delta P = -q/2$.
  It trivially maps to our three-dimensional system, the point being 
that  different bulk polarisation reference definitions can be used,
requiring a consistent redefinition of the interface charge.
  The original paper \cite{Ohtomo2004} and most that followed used 
the split-charge trick on the formal charges of the layers.

  Quantum-mechanical electrons pose additional difficulties to the definition
of the bulk polarisation of a solid~\cite{Martin74}, given their spread charge density: 
When displacing the unit cell, there is a continuous unphysical change of $P$ if
defined as above, instead of the jumps of quanta discussed above.
  The subtleties of a meaningful definition are illustrated for a chain in the last
section of this paper.
  For the point at hand, however, it is sufficient to know that towards calculating 
$P$ the modern theory of polarisation~\cite{KingSmith1993,Resta1994} allows 
to map the electronic wave functions in a material onto a classical system of point 
charges, using the centres of charge of the Wannier functions associated 
to the valence bands~\cite{Marzari1997}.
  Taking the ionic cores and the Wannier centres in a unit cell as point 
charges, the polarisation of any material can be defined as above, running 
onto the same arbitrariness.
  Equally, the inversion symmetry allows taking both bulk phases of
LAO and STO with $\vec P = 0$.
  The counting of charges (core charges and Wannier charges), gives the 
same 0.5$e$/f.u. (for a figure illustrating it for the LAO/STO interface, 
see Figure 1 of Ref.~\cite{Stengel2009}).

\subsection{Symmetry considerations}

  The well defined polarisation reference for bulk centrosymmetric solids
has been explicitly used above.\footnote{Note that, 
although we make the case for inversion, the only invariance needed in 
this discussion is that of any symmetry transformation that transforms 
$z$ into $-z$. See Ref.~\cite{Vanderbilt1993} for a more general discussion.}
  It is more general than it seems.
  The vast majority of symmetry breaking displacive instabilities in 
perovskites either remain centrosymmetric or become ferroelectric,
both allowing for a good definition of a polarisation zero.

  The situation is different for non-centrosymmetric
non-switchable solids. 
  In such a case, the definition of an effective polarisation 
taking as reference the one found in the unperturbed solid remains 
quite arbitrary.
  It actually changes with temperature (pyroelectricity).
  The zero temperature limit can then be used (as defined 
in~\cite{Stengel2009}), but it still remains unsatisfactorily arbitrary.
  The key here is that for any interface involving such a solid,
any unit cell definition giving zero polarisation would give an
arbitrary fractional charge at the interface, apparently calling
for ionicity/covalency considerations at the interface 
\cite{Frensley1977,Baraff1977}.
  The muddled situation generated by the lack of symmetry in these 
materials clears below. 
   
\section{Formal versus effective polarisation}

  Stengel and Vanderbilt~\cite{Stengel2009} propose a more elegant 
alternative view on this problem based on the so-called interface 
theorem~\cite{Vanderbilt1993}.
  They come from the fact that the charge density entering Gauss's law
is that of free charges while ``bound" charges are described by the 
polarisation term in the displacement field. 
  The net compositional charge discussed here is neither.
  A more natural solution appears when 
considering the {\it formal} values of polarisation of both materials 
instead of their {\it effective} values~\cite{Stengel2009}.
  The former is as obtained by direct application of the modern theory 
of polarisation mentioned above~\cite{KingSmith1993} (e.g. considering 
the ionic cores and the centres of charge of Wannier functions as point 
charges \cite{Marzari1997}), which offers a well defined zero reference. 
  The difference between the two definitions gives precisely 0.5$e$/f.u.: 
  The formal $P_z$ for STO is zero, while for LAO it is 0.5$e$/f.u. 
(see Ref.~\cite{Stengel2009}). 
  With these values for the respective bulk polarisations, 
the net charge at the interface is now zero (see first case in Figure 1b).

\subsection{Polarisation quanta and symmetry}

  It is interesting to see how these two polarisation values come about.
  In a system of discrete point charges, the polarisation is defined up to 
polarisation quanta: finite jumps of the polarisation
when displacing the bulk unit cell along a crystallographic direction.
  A jump occurs when the edge of the unit cell reaches one of the point 
charges,  which disappears there and reappears on the other side
of the cell, and thus the dipole moment for the unit cell
is changed by $qL$, the charge of the point-like particle that flipped, 
times the length of the cell in the direction of the flip.
  The polarisation changes by $qL/\Omega$, 
where $\Omega$ is the unit cell volume, or $q/A$, being $A$ the
area of the unit cell normal to the flip direction.
  With $e$ as the charge quantum, the quantum of polarisation in a 
given direction $P_0$ is $1e/A$, and in our case $P_0 =1e$/f.u.

  In these flips the system does not change at all, only our description of it.
  Such jumps in the value of $P$ have thus no physical meaning.
  A meaningful polarisation value is thus modulo polarisation quanta.
  In other words, the whole set $\{P+nP_0\}$ ($n$ being any integer)
corresponds to one only value. 
  In our problem, only two $P_z$ values are allowed if the 
material is centrosymmetric:  $P_z = 0$ or $P_z=P_0/2$, corresponding
to the sets  $P_z = \{ ... -2P_0, -P_0, 0, P_0, 2P_0, ...\}$ and 
$P_z = \{ ... -3P_0/2, -P_0/2, P_0/2, 3P_0/2, ...\}$, respectively, since 
these are the only such sets that are invariant under change of sign.
  This is the key difference with our discussion in the previous section, 
in which the {\it effective} polarisation of any centrosymmetric system
was defined to be always zero, instead of zero or $P_0/2$.

\section{Further implications}

\subsection{Non-centrosymmetric non-ferroelectric materials}
  
  The proposal of using formal values of the polarisation is even more
interesting if the materials are neither centrosymmetric nor ferroelectric,
as is the case of III-V semiconductors as GaAs or GaN, as they
interface with a non-polar one as Si (take (111) interfaces for 
diamond/sphalerite structures, (001) for wurtzite).
  The trick used in the previous section (take zero-polarisation materials
at each side), is quite contrived since, for the polar material, one has to 
split a point charge into two arbitrary bits to be put at each side of the cell 
in order to get $P=0$.
   It leaves some inconvenient non-zero value of net interface charge
(one of the bits, actually equal to the formal value of the polarisation).
  The consideration (and calculation) of the formal polarisation as such, 
however, simplifies matters, implying zero compositional charge at the 
interface.
  This is concluded following the same procedure as before:
  ($i$) take the system as made of point charges associated to the ionic 
cores and the Wannier centres.
  ($ii$) define your unit cells such that all charges are accounted for at
the interface.
  ($iii$) Consider the corresponding formal polarisation values arising 
from it; and  ($iv$) count the extra charges at the interface.
  The apparent ``covalency" issue that appeared at the interface reduces
to the formal value of the polarisation in the bulk materials.
  The bond covalency/ionicity factors introduced in the past 
\cite{Frensley1977} or other charge density, or charge-density ``topology" 
considerations (see them reviewed in \cite{Goniakowski2008}) are 
not addressing the problem from the right angle.

\subsection{Polarisation quanta and relative polarisation}  

  Going back to centrosymmetric perovskites, consider now the
interface between KTaO$_3$ and LaAlO$_3$ (KTO/LAO).
  The formal polarisation for both materials taken separately 
is $P_0/2$.
  Taking these values and assuming zero compositional charge would
be wrong, however.
  This can be seen if computing the compositional charge as in section 3.
  In this case, while in LAO the AlO$_2$ planes are have a $-1$ charge, and
the LaO planes are $+1$, in KTaO it is the other way around: TaO$_2$ planes 
are $+1$, while KO planes are $-1$.
  The two possible interfaces would then be TaO$_2$/LaO, or
KO/AlO$_2$, which gives net compositional charges of $-e$/f.u.
and $+e$/f.u., respectively, which coincide with $\pm P_0$.
  A quantum of polarisation can thus have physical meaning. 
  It does when determining the relative polarisation of the two materials.
  One can define the formal polarisation of 
one material with an arbitrary number of added polarisation quanta, but the 
condition of zero compositional charge at the interface completely defines 
the polarisation of the material at the other side of the interface.
  In this case $\sigma_{\rm comp}=0$ if taking $P=P_0/2$ for one
and $P=-P_0/2$ for the other.

\subsection{Internal rearrangements versus extrinsic processes}

  This discussion has implications on the electrostatics
of chemically altered interfaces.
  In particular, within a certain width of a given interface,
whatever internal rearrangement of atoms has no effect on the
electrostatics at larger length scales.
  These rearrangements can affect interface dipoles, but not 
the difference in outgoing electric fields at either side of
the interface, $\Delta {\cal E}_z$.
  The fields at either side can change, for instance, by altering an 
interface dipole under fixed-bias boundary conditions.
  This is an indirect global effect, however, whereby a constant 
field is added to the whole system
leaving $\Delta {\cal E}_z$ invariant.
  This means that a diffusive disordering of the cations has no
possibility of directly affecting the electrostatics, if the chemistry 
(in terms of a clean valence band and gap) is preserved, and
the length scale of the inter-diffusion is smaller than the thickness
of the film.

  The electrostatics is affected by extrinsic processes such as receiving 
free carriers, altering the stoichiometry at growth, or redox processes 
at the interface. 
  They can also combine, like a remote redox process 
at the surface of a film that sends free carriers to a buried interface. 
  An important redox process in these materials is the formation of
O vacancies, whereby the pristine solid is reduced leaving 
two electron carriers.
  The vacancy becomes a centre with an effective $Z=+2$ by
the same arguments as in semiconductor doping (section 2):
a double donor.
  In the bulk the two free electrons bind to the vacancy while
in heterostructures they may find more stable residence elsewhere,
in perfect analogy to what happens in semiconductor
heterostructures.
  O vacancies on the surface of LAO may see their electrons 
transferred to the two-dimensional states at the buried STO/LAO
interface \cite{Cen2008,Bristowe2010a}

  As a final remark let us note that although the examples illustrating
this discussion have been drawn from band insulators (doubly occupied
Wannier functions), the arguments are more general
and include Mott insulators, as long as there is a well defined
gap (think in terms of spin-polarised wave-functions and their
corresponding, singly occupied Wannier functions).
  This is so for interfaces but also for point defects.
  In fact, the point made in Ref.~\cite{Raebiger2008} on distinguishing
valence from charge density effects in transition metal oxides and
impurities in them is easily understood in terms of the arguments 
presented here.



\section{Polarisation for quantum electrons: linear chain model}

  Some insight on how this works for a quantum system can be gained by 
studying a simple model that is analytically solvable, a diatomic linear chain.
  Consider two atoms per unit cell (A,B) at the positions $\pm \frac{1}{2}(\frac{1}{2} a + u)$, 
where $a$ is the lattice constant, and $u$ a parameter that measures the 
displacement from equidistant.
  Note that if $u = 0$ the system has inversion symmetry about either of the 
atoms; for a small displacement, one expects a polarisation linear in $u$, 
conventionally written as $P = e Z_T^* u$ (we are considering here the
{\it effective} polarisation).
  Notice that $u$ is just the amplitude of a $q=0$ {\em optical } phonon.
  Here $Z_T^*$ is the Born effective charge, which is directly measured 
in the splitting of a LO and TO phonon.
  Notice of course that should the atoms be chemically identical, viz $A=B$, 
then there is an inversion centre midway between each pair of atoms, even 
if $u$ is non-zero.

  Take the case of two electrons per unit cell. 
  A filled Brillouin zone suggests that a good approximation to calculate $P$ 
will be just the nearly free electron approximation with a single band gap.
  This is achieved by just keeping a single Fourier component of the 
backscattering by the atoms, namely that 
$V(x) = V e^{i 2 \pi x / a} + V^* e^{-i 2 \pi x / a}$.
  A bit of algebra relates $V$ to its atomic form factors $V_A$ and $V_B$:
\begin{equation}
V = (V_A + V_B) \sin \left ( \frac{\pi u }{a} \right) + 
       i (V_A - V_B) \cos \left ( \frac{\pi u }{a} \right) .
\end{equation}
  One may calculate~\cite{Littlewood80} $P$ as a function of $u$, and the result in 
the linear regime is easily stated (and exact for the model above, because it 
can be derived from a sum rule), viz.
\begin{equation}
\label{eq:Zeff}
Z_T^* = \frac{1}{2} (Z_A - Z_B) +  \frac {(V_A + V_B)}{(V_A - V_B)}
\end{equation}
where the first term comes from the fixed ion cores of charges $Z_{A,B}$, and the 
second is the electronic component. Notice that the effective charge becomes 
very large if $V_A-V_B$ is small - this remarkable result arises because if the 
ionicity is nearly vanishing, the effect of the displacement $u$ is to produce 
a rapid gap-opening, and a considerable displacement of electronic charge into 
the dimerised bond.
  Because our centre of symmetry was originally centred on an atom, this 
generates a much larger dipole than a rigid displacement of charge.
  Such an effect is well known in narrow gap semiconductors such as GeTe 
or Bi$_2$Te$_3$, where measured values of $Z_T^*$ can exceed 10.


\begin{figure}
\begin{center}
\includegraphics[width=2.5in]{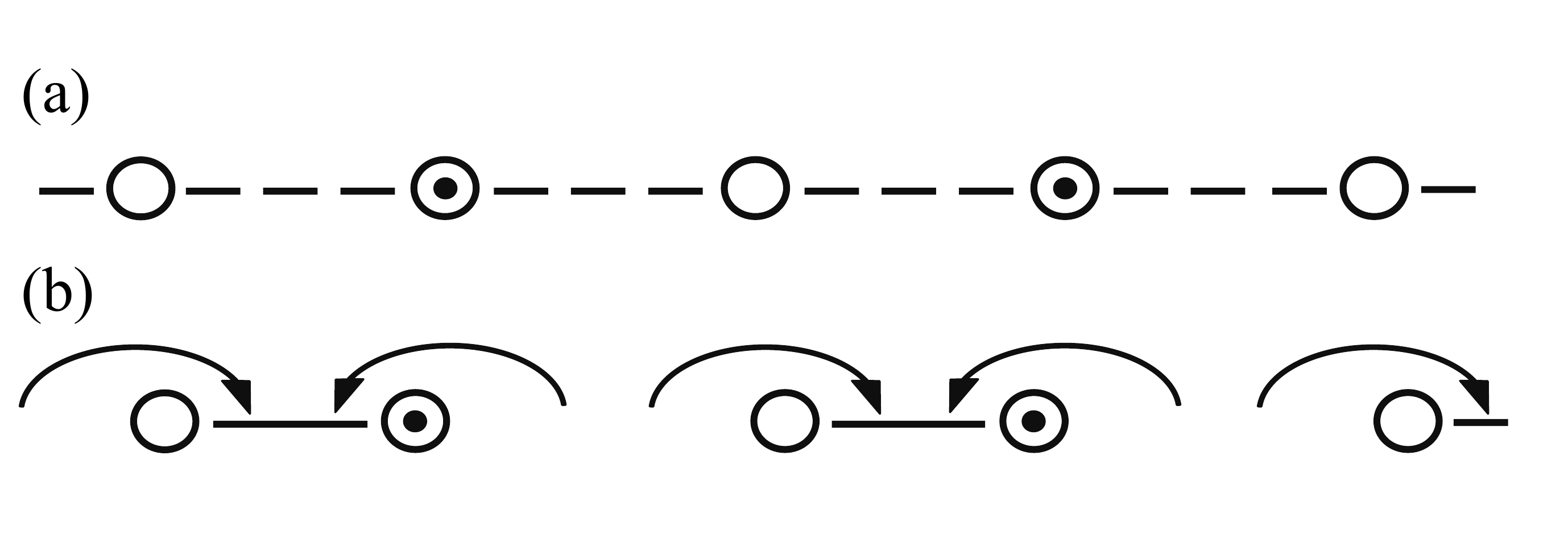} \\
\includegraphics[width=3.0in]{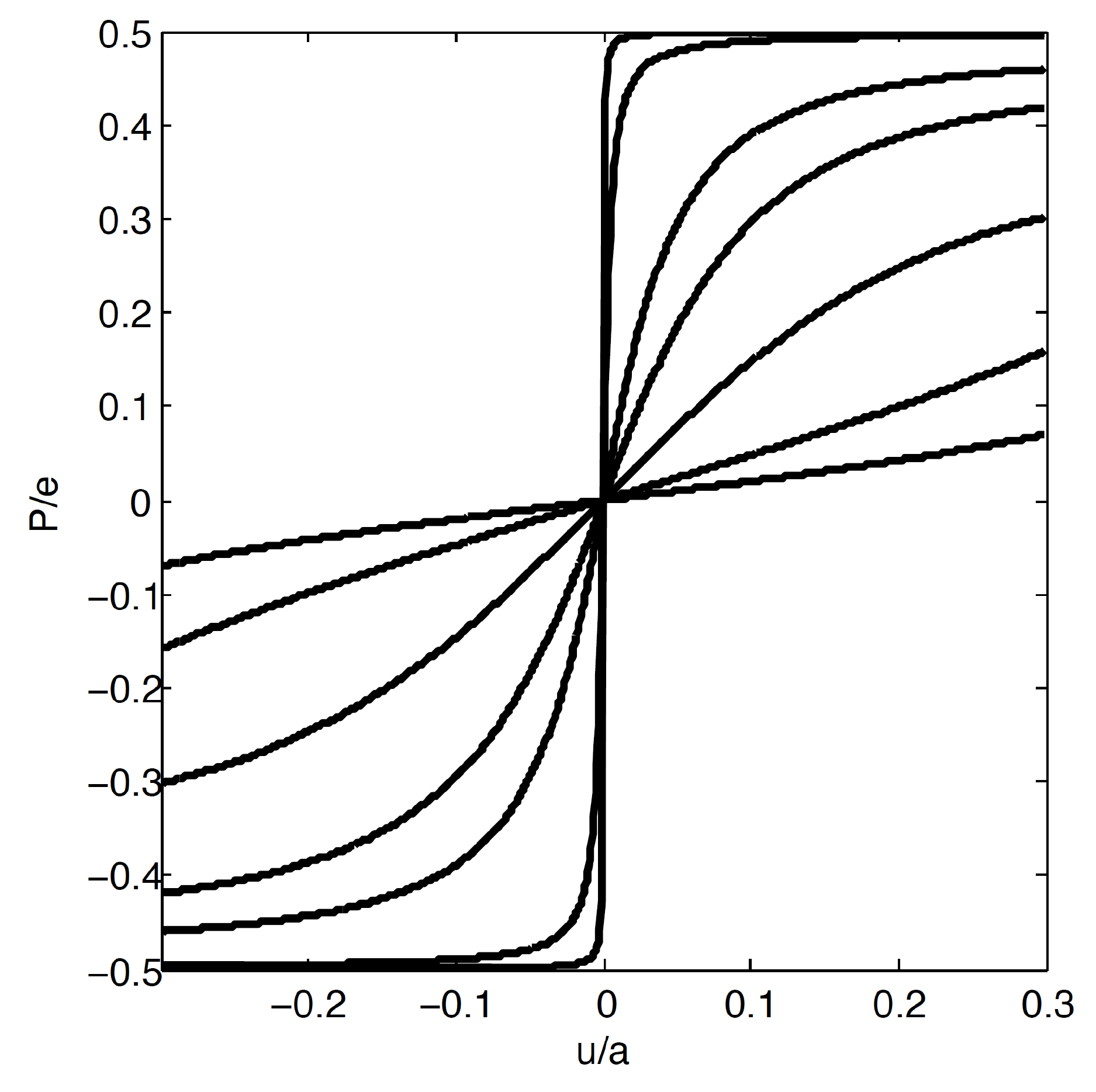}
\end{center}
\caption{In the undistorted structure (a) each covalent bond is partially completed. 
  In the distorted structure (b) the electrons migrate to complete the covalent bond 
formed between the closer atoms.
  With reference to a symmetry point at the position of an undisplaced atom, this 
generates the polarisation, plotted as a function of the optical phonon 
displacement $u$.
  Different curves correspond to the parameters 
$\frac {(V_A + V_B)}{(V_A - V_B)}$ = (0.2, 0.5, 2, 5, 10, 100, 1000,)}
\end{figure}

  Equation \ref{eq:Zeff} {\it diverges} if the atoms are identical --- yet the identical 
atom situation is inversion-symmetric and the bulk polarisation should vanish.
  The paradox is cleared if we look at the full non-linear form for $P(u)$ in 
Figure 2, plotted for different ratios of the symmetric to antisymmetric potentials.
  Notice that $P(u)$ saturates with growing displacement at a value of $0.5 e$ 
- there are only so many electrons to move around.
  Notice also that as $V_A - V_B \rightarrow 0$, the polarisation converges to a 
step function, with a jump $\Delta P = e$.
  Our ``error'' was  that in the limit of identical atoms, we computed the polarisation 
with reference to the wrong centre of symmetry (now centred between the atoms, 
not on either of them).
  Furthermore, for identical atoms and zero displacement the band-gap vanishes: 
the macroscopic electric field is annulled at no energy cost by taking a single 
electron from one end of the chain to the other, just as we saw in the classical 
description above.
  This is the physical content of the modern Berry phase description of polarisation.
  It connects easily with the previous discussion for the effective polarisation.
  For the {\it formal} polarisation, Fig.~2 displaces up half a quantum, giving a 
zero value for the purely covalent case (just think of the two-electron Wannier function
centred exactly in the middle of the dimer, and consider all the charges not split
in the unit cell). 
  In the ionic case the formal $P$ tends to 0.5$e$ for small $u$, i.e., half a quantum, 
as in LAO.

  In conclusion we have reviewed the issue of the net charge at
interfaces between insulators. 
  In an attempt to dispel myths of such charges being affected by 
covalency and related charge density effects, we have presented a case for
the counting of discrete electrons and core ion charges or equivalently the use of the 
formal polarisation value. 
  The arguments are illustrated for the popular LAO/STO interface 
and additionally for non-centrosymmetric materials and
for interfaces for which the charge mismatch is an integer number of polarisation quanta. 

\ack
  We wish to thank H Y Hwang, S A T Redfern, P Ordej\'on and
D Vanderbilt for useful discussions.
  This work has been partly funded by UK's EPSRC.

\section*{References}



\providecommand{\newblock}{}

\end{document}